\documentclass[conference]{IEEEtran}
\IEEEoverridecommandlockouts
\usepackage{cite}
\usepackage{amsmath,amssymb,amsfonts}
\usepackage{algorithmic}
\usepackage{graphicx}
\usepackage{textcomp}
\usepackage{xcolor}
\usepackage{url}
\def\BibTeX{{\rm B\kern-.05em{\sc i\kern-.025em b}\kern-.08em
    T\kern-.1667em\lower.7ex\hbox{E}\kern-.125emX}}
\begin{document}

\title{Compression of Solar Spectroscopic Observations: \\a Case Study of Mg\,II\,k Spectral Line Profiles Observed by NASA's IRIS Satellite\\
\thanks{VMS is supported by NSF FDSS grant 1936361 and NSF grant 1835958. EI is supported by RSF grant 20-72-00106. The research is partially supported by NASA grants NNX12AD05A, 80NSSC20K0302, and NSF EarthCube grant 1639683.}}

\author{\IEEEauthorblockN{Viacheslav M Sadykov}
\IEEEauthorblockA{\textit{Physics \& Astronomy Department} \\
\textit{Georgia State University}\\
Atlanta, GA, USA \\
vsadykov@gsu.edu}
\and

\IEEEauthorblockN{Irina N Kitiashvili}
\IEEEauthorblockA{\textit{NASA Supercomputing Division} \\
\textit{NASA Ames Research Center}\\
Moffett Field, CA, USA \\
irina.n.kitiashvili@nasa.gov}
\and

\IEEEauthorblockN{Alberto Sainz Dalda}
\IEEEauthorblockA{\textit{Solar \& Astrophysics Laboratory} \\
\textit{Lockheed Martin}\\
Palo Alto, CA, USA \\
asainz.solarphysics@gmail.com}
\and

\IEEEauthorblockN{Vincent Oria}
\IEEEauthorblockA{\textit{Computer Science Department} \\
\textit{New Jersey Institute of Technology}\\
Newark, NJ, USA \\
vincent.oria@njit.edu}
\and

\IEEEauthorblockN{Alexander G Kosovichev}
\IEEEauthorblockA{\textit{Physics Department} \\
\textit{New Jersey Institute of Technology}\\
Newark, NJ, USA \\
alexander.g.kosovichev@njit.edu}
\and

\IEEEauthorblockN{Egor Illarionov}
\IEEEauthorblockA{\textit{Department of Mechanics and Mathematics} \\
\textit{Moscow State University}\\
Moscow, Russia \\
egor.mypost@gmail.com}

}

\maketitle

\begin{abstract}
In this study we extract the deep features and investigate the compression of the Mg\,II\,k spectral line profiles observed in quiet Sun regions by NASA's IRIS satellite. The data set of line profiles used for the analysis was obtained on April 20th, 2020, at the center of the solar disc, and contains almost 300,000 individual Mg\,II\,k line profiles after data cleaning. The data are separated into train and test subsets. The train subset was used to train the autoencoder of the varying embedding layer size. The early stopping criterion was implemented on the test subset to prevent the model from overfitting. Our results indicate that it is possible to compress the spectral line profiles more than 27 times (which corresponds to the reduction of the data dimensionality from 110 to 4) while having a 4\,DN (Data Number) average reconstruction error, which is comparable to the variations in the line continuum. The mean squared error and the reconstruction error of even statistical moments sharply decrease when the dimensionality of the embedding layer increases from 1 to 4 and almost stop decreasing for higher numbers. The observed occasional improvements in training for values higher than 4 indicate that a better compact embedding may potentially be obtained if other training strategies and longer training times are used. The features learned for the critical four-dimensional case can be interpreted. In particular, three of these four features mainly control the line width, line asymmetry, and line dip formation respectively. The presented results are the first attempt to obtain a compact embedding for spectroscopic line profiles and confirm the value of this approach, in particular for feature extraction, data compression, and denoising.
\end{abstract}


\section{Introduction}
\label{section:introduction}

    Solar spectral lines in the infrared, visible, and ultraviolet ranges are powerful diagnostics for the solar atmosphere. Many of the spectral lines originate in the solar atmosphere (the photosphere, chromosphere, and the chromosphere-corona transition region) and reveal the state and processes in this complex and highly dynamic region. The line profiles reflect the distribution and dynamics of physical parameters (density, temperature, macroscopic and turbulent velocities, etc.) as functions of height in the solar atmosphere. Correspondingly, analysis and interpretation of solar spectral lines becomes an extremely important task for understanding physical processes in the solar atmosphere.
    
    With the launch of NASA's Interface Region Imaging Spectrograph (IRIS, \cite{DePontieu2014}), the diagnostics of the solar chromosphere and transition region was significantly advanced. IRIS observes a variety of lines formed in these regions (Mg\,II\,k\&h 2796\,\AA\, and 2803\,\AA\,, C\,II 1334\,\AA\, and 1335\,\AA\,, Si\,IV 1394\,\AA\, and 1403\,\AA), each sensitive to certain ranges of temperatures and, correspondingly, to certain atmospheric heights. In this work, our analysis will be focused on the Mg\,II\,k line profiles. It was previously demonstrated that Mg\,II\,k\&\,h lines may serve as diagnostic tools for probing parameters of the solar atmosphere in the upper chromosphere \cite{Leenaarts2013,Sadykov2021}. In addition to a variety of physics-based modeling, analysis, and inversion techniques, there have been several attempts to apply machine learning for enhancing the physics knowledge obtained from Mg\,II\,k line profiles. For example, more than 50000 {\it representative profiles} (RP) of the Mg\,II\,k\&h lines were computed by applying clustering algorithms in an assortment of IRIS observations \cite{SainzDalda2019}. The spectra were inverted using a physics-based approach, and the inversion results were used to train a deep learning model emulating the physics-based inversions. Thus, the physical mode can be obtained in a few minutes using a laptop or a desktop machine, which is $\sim$10$^5$-10$^6$ times faster than the classic physics-based inversions. The typical Mg\,II\,k\&h line profiles observed during solar flares were identified using unsupervised clustering algorithms \cite{Panos2018} and linked the appearance of certain types of clusters to specific physical processes occurring during solar flares (electron beam injection properties and flare ribbon propagation).
    
    While the application of machine learning in solar physics is an expanding field \cite{Nita2020}, many questions are still not addressed. One of them is the question of compact embedding (extraction of the deep features) for spectral line profiles, which became of interest to the community just recently \cite{Panos2021}. Obtaining low-dimensional representations of spectra may potentially help address noise reduction, feature selection, and data compression and transmission problems. This manuscript describes an attempt to apply compact embedding analysis to spectral lines. The experiment shows that one needs only 4 dimensions to represent the studied data set. Section~\ref{section:data} describes the data set of Mg\,II\,k line profile observations, data cleaning, and normalization procedures. Section~\ref{section:autoencoder} illustrates the architecture of a neural network used to obtain a compact embedding of the line profiles. Analysis of reconstruction errors for different embedding layer dimensionalities is discussed in Section~\ref{section:results}. An interpretation of the results is summarized in Section~\ref{section:discussion} followed by conclusions in Section~\ref{section:conclusion}.

\section{Data}
\label{section:data}

\begin{figure*}[t]
    \centerline{\includegraphics[width=1.0\linewidth]{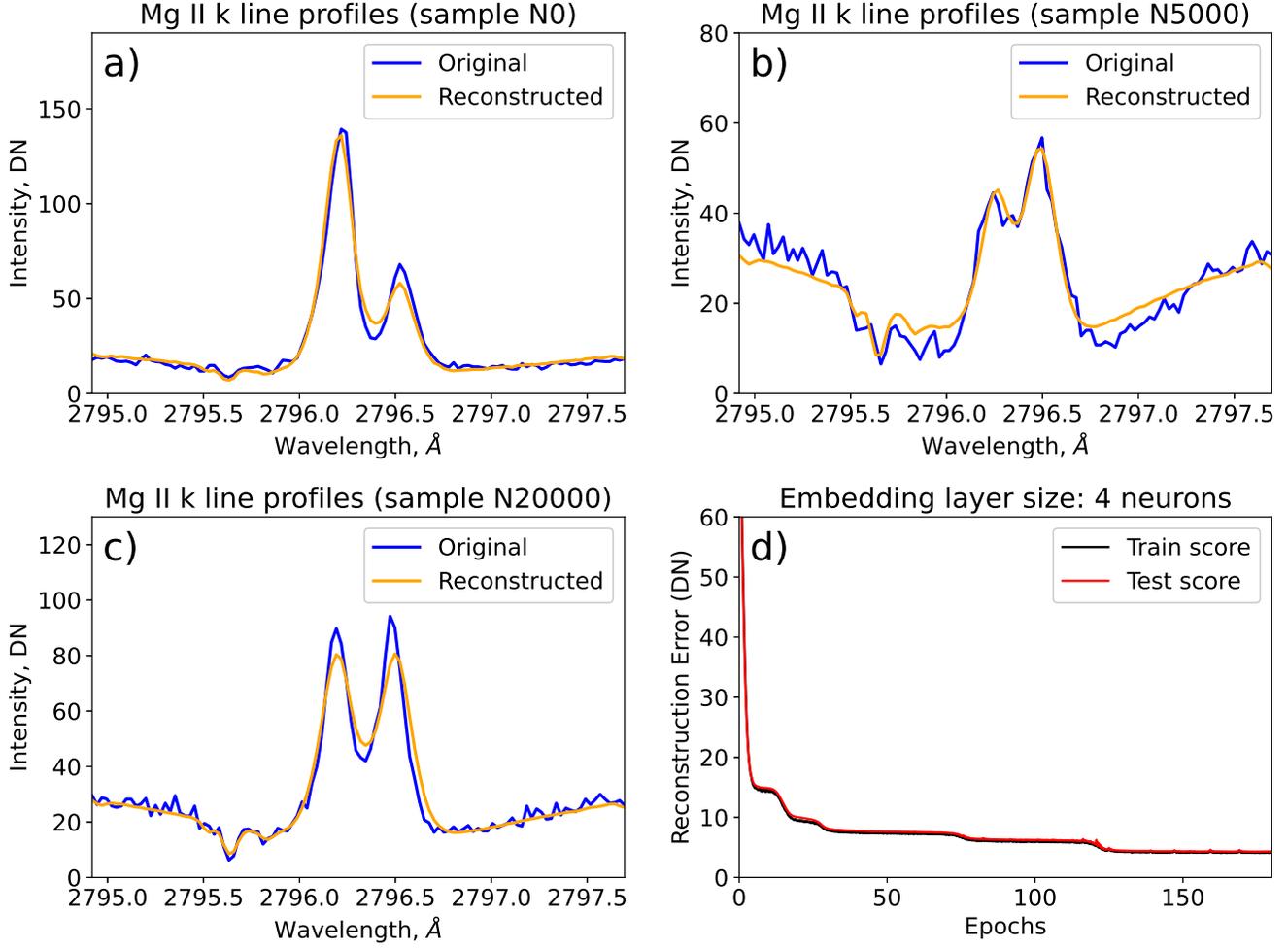}}
    \caption{Examples of the original (blue) and reconstructed (orange) Mg\,II\,k line profiles from the test subset for the case of an embedding layer size of 4 neurons. Lower right panel illustrates the corresponding autoencoder training process (mean squared error, MSE, as a function of the epoch number).}
    \label{figure:examples}
    \end{figure*}

    We utilize IRIS observations of the quiet Sun taken on April 20, 2020 from 08:32:00 UT - 09:56:00 UT at the center of the solar disk. The observations were made in the sit-and-stare mode (the mode when the slit of the spectrograph is always in one position and ``tracks'' the solar features). During these observations, IRIS obtained more than 300,000 individual profiles of Mg\,II\,k lines sampled from the quiet Sun at different time moments and positions along the slit. Some examples of the obtained line profiles are presented in Figure~\ref{figure:examples}. One can notice that the Mg\,II\,k spectral line has a complex shape~-- it typically has a double-peak structure with a central-reversal signature if observed in quiet Sun regions. Both the locations of the peaks and the central reversal represent valuable features for diagnostics of the solar atmosphere \cite{Leenaarts2013}. It is important to note that we consider only the Mg\,II\,k line-related part of the observed spectra; it consists of intensity measurements at 110 wavelengths covering the 2794.9\,\AA\, -- 2797.7\,\AA\, spectral range; each wavelength represents a separate dimension.
    
    The pool of obtained line profiles was cleaned. Any intensity points beyond the 2794.9\,\AA\, -- 2797.7\,\AA\, spectral range were removed from the analysis. The line profiles with spikes in the data (defined here as sudden increases of the intensity of 1500\,DN or more at a particular wavelength) and measurements with negative intensity values were removed from consideration. The 1500\,DN threshold was defined as a minimum possible threshold that keeps all the line profiles not affected by spikes in the data set. The resulting 290940 Mg\,II\,k line profiles were all normalized to the same number, which is the maximum peak intensity observed across the line profiles. The first 200000 normalized line profiles were assigned to the train data subset and the remaining 90940 profiles to the test data subset, giving an approximate partitioning of 70\% to 30\%.

\section{Autoencoder Configuration and Training}
\label{section:autoencoder}

    For compression of spectroscopic line profiles we have utilized an autoencoder-type neural network architecture. The autoencoder is represented by a fully connected neural network with the numbers of neurons in the subsequent layers set at 110 - 64 - 32 - 16 - $n_{emb}$ - 16 - 32 - 64 - 110. The network was implemented in Python PyTorch \cite{Paszke2017}. Here $n_{emb}$ represents the dimensionality of the embedding layer and is varied from 1 to 15. The Rectified Linear Unit (ReLU) activation function was used for all neurons, and the Mean Squared Error (MSE) was utilized as a loss function. A batch size of 20,000 samples was used. Two network optimizers, ``adam'' \cite{Kingma2015} and stochastic gradient descent, and three learning rates (0.001, 0.00033, 0.0001) were utilized to progress the training. We found that introducing additional fully-connected layers into the network or increasing the number of neurons in the layers does not change the conclusions of this work.
    
    To ensure the completeness of the training process and prevent the network from overfitting, we have implemented the early stopping criterion. Specifically, we assumed that the training process of the network for the current setting is complete if the MSE on the test subset is not improving for more than 0.5\% for two consequent training epochs. The strategy was sequentially applied for each optimizer and learning rate combination (with the learning rates arranged in decreasing order). The entire training process was repeated five times for each $n_{emb}$ to check how variations in the initial neuron weight distribution affect the training.
    
    Figure~\ref{figure:examples} illustrates the reconstructed Mg\,II\,k line profiles and the corresponding training progress for $n_{emb}$=4. The network captures the main line profile features (the central-reversal, the peak asymmetry, and the line continuum behavior and features therein) relatively well, at least from a qualitative perspective.

\section{Results of Reconstruction}
\label{section:results}

    \begin{figure*}[t]
    \centerline{\includegraphics[width=1.0\linewidth]{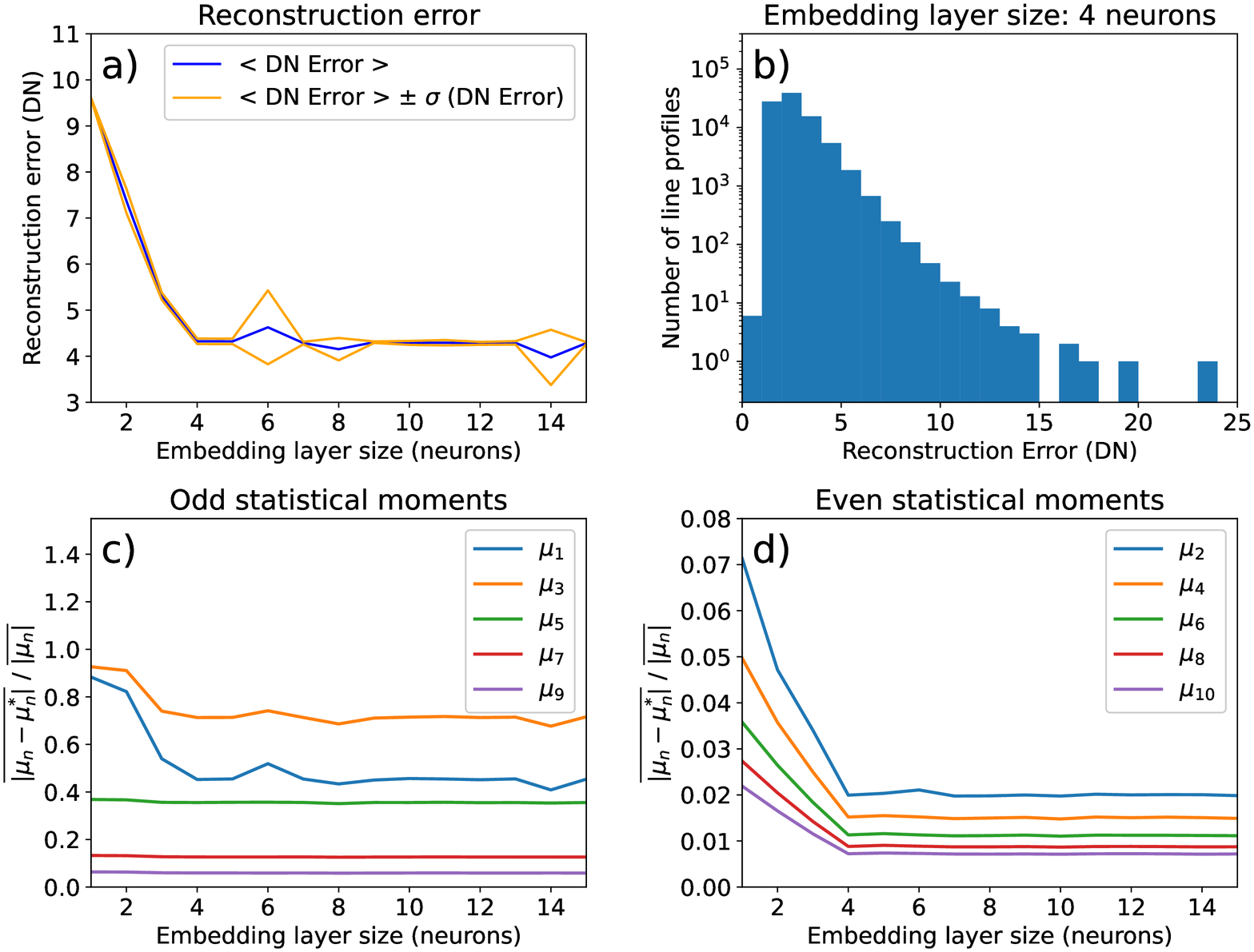}}
    \caption{Average error of the line profile reconstruction as a function of the embedding layer dimensionality. The upper-left panel shows the mean squared error (MSE) of the line profile intensity reconstruction, and the corresponding histogram of the distribution of such errors is presented in the upper right panel. The lower-left panel presents the relative mean error of the reconstruction of the odd statistical moments and the lower-right panel that of the even moments.}
    \label{figure:dimvars}
    \end{figure*}

    The average MSE as a function of $n_{emb}$ (in the original intensity units of data numbers, DN) is presented in Figure~\ref{figure:dimvars}\textit{a}. One can notice that the MSE decreases sharply for $n_{emb} \leq{} 4$ and almost stops decreasing when $n_{emb} > 4$. One can state that there is no further progress in line compression for $n_{emb} > 4$ for the current training setting and autoencoder architecture. The distribution of MSE for the case of $n_{emb} = 4$ is presented in the upper-right panel of the same figure. For the majority of the line profiles, the average deviation of the reconstructed intensity from the original values was around 3\,DN, which is comparable with the intensity variations of the line continuum signal.
    
    In addition to the MSE, we also consider the reconstruction of the statistical moments of the spectral lines. In this study, the n-th statistical moment of the line profile is defined as:
    \begin{equation}
        \mu{}_{n} = \left( \dfrac{ \int_{-\infty}^{\infty}(\lambda{}-\lambda{}_{0})^{n}I(\lambda{})d\lambda{} }
                    { \int_{-\infty}^{\infty}I(\lambda{})d\lambda{} } \right)^{1/n},
    \end{equation}
    \begin{equation}
        \lambda{}_{0} = \dfrac{ \int_{-\infty}^{\infty}\lambda{}I(\lambda{})d\lambda{} }
                    { \int_{-\infty}^{\infty}I(\lambda{})d\lambda{} },
    \end{equation}
    Where $\lambda{}$ is a wavelength point and $I(\lambda)$ is the intensity at that wavelength. Typically, the first statistical moment serves as a proxy of for line Doppler shift, the second one as a proxy for the line width, etc. These characteristics are used in spectroscopic studies \cite{Sadykov2015}. Therefore, it is important to analyze the accuracy of the reconstruction for these moments. In this work, the statistical moments are computed only for the 2795.68\,$\AA$-2796.96\,$\AA$ spectral range, which covers the Mg\,II\,k line but removes most of the contribution from the continuum.
    
    \begin{figure*}[t]
    \centerline{\includegraphics[width=1.0\linewidth]{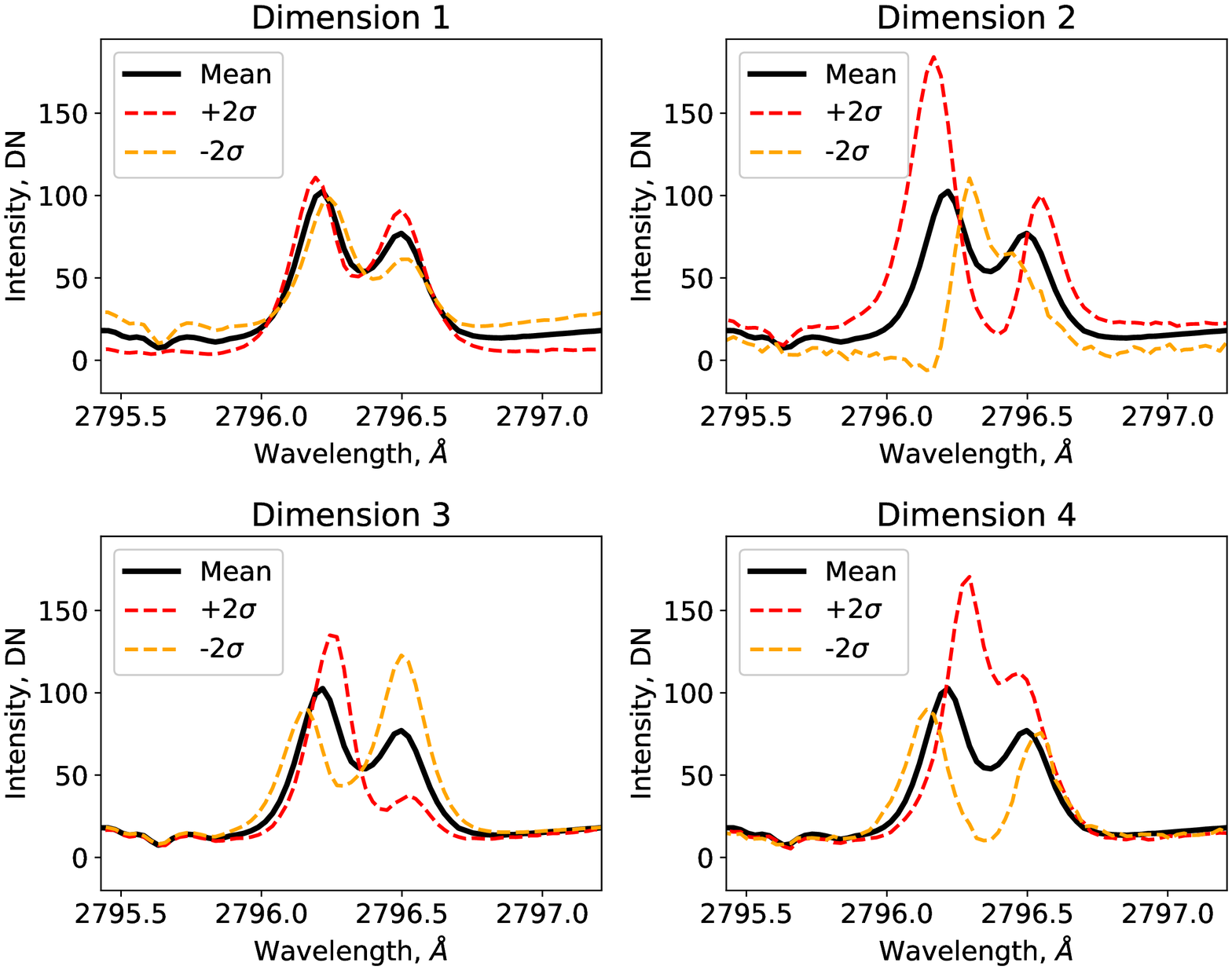}}
    \caption{Variations of the decoded line profiles as results of perturbations of the embedding space. Black curves correspond to the line profiles reconstructed from the average embedding parameters. Red and orange curves correspond to the situation when one of the embedding parameters is one standard deviation larger or smaller than the average. The perturbed parameter is indicated in the plot title.}
    \label{figure:diminterp}
    \end{figure*}
    
    The errors of reconstruction of the statistical moments are presented in Figure~\ref{figure:dimvars}\textit{c,d}. As one can see in the lower-right panel, the even statistical moments experience the same behavior as MSE does: the error of reconstruction of these line moments improves for $n_{emb} \leq{} 4$ and stops decreasing for $n_{emb} > 4$. The considered moments have an average relative error of the reconstruction of $\leq{}$\,2\%, and the error decreases for higher moments.
    
    Surprisingly, the situation is different for the odd line profile moments. First, the odd (asymmetric) line moments experience significantly higher average errors, although, starting from $\mu{}_{5}$, the reconstruction error is below 40\% on average (Fig.~\ref{figure:dimvars}\textit{c}). Second, the reconstruction errors for the odd statistical moments does not decrease with increasing $n_{emb}$ as strongly as was observed for the even components (except the $\mu{}_{1}$ case, which also had a $\sim{}$\,50\% reconstruction error even for high $n_{emb}$).

\section{Discussion. Special case of $n_{emb}$=4.}
\label{section:discussion}

    As mentioned in Section~\ref{section:introduction}, finding a compact embedding of spectroscopic line profiles has several important applications. In this work we consider, in particular, compression of the 110-dimensional Mg\,II\,k line profile into the $n_{emb} = [1; 15]$ space and analyze the resulting reconstruction error. One of the key results of such analysis is that the reconstruction error (in terms of MSE or the even statistical moments) does not decrease for $n_{emb} > 4$ (Fig.~\ref{figure:dimvars}). Such behavior may have two explanations. The first is that the entire physics of the line formation may be expressed in four free parameters (e.g., the temperature, density, the macroscopic and turbulent velocities in case of the homogeneous thermal media). Another possible explanation is that different neural network architectures and training strategies might achieve better performance of the autoencoder for $n_{emb} > 4$. Because there was a specific architecture of the network and a specific training strategy considered, this is a very probable scenario. One also can see that, for $n_{emb} = 8$ and $n_{emb} = 14$, there were breakthroughs for some particular trials. Specifically, one of the trials demonstrated a $\sim$20\% improvement of MSE for $n_{emb} = 8$, and one of the trials demonstrated a 55\% improvement of MSE for $n_{emb} = 14$. As one can see, it may possible to have better compressions of line profiles, but that will require more computational resources.

    To understand better the meaning of the parameters that the autoencoder learned for the $n_{emb} = 4$ case, we conduct the following experiment on the embedding space. We calculate the means and the standard deviations for all parameters (dimensions) of the embedding space for the line profiles in the test data set. Next, we vary one of the four dimensions in the range of the mean $\pm$ two standard deviations, while keeping the other three parameters fixed and equal to their mean values. The result of the experiment is presented in Figure~\ref{figure:diminterp}. 
    
    As one can see, sometimes the meaning of the dimensions learned by the autoencoder could be intuitively interpreted. For example, dimension\,\#3 is mostly responsible for the line profile asymmetry, dimension\,\#4 is mostly responsible for the central-reversal feature depth, and dimension\,\#2 significantly affects the line profile width. The meaning of dimension\,\#1 is less clear, but it seems to control the line profile continuum with some effect on the line profile itself. Overall, it is very intriguing that the features learned by the network for the $n_{emb} = 4$ case can be interpreted and explained, to a certain degree, relatively simply. We would like to note that the optimal embedding layer dimensionality of $n_{emb} = 4$ and the interpretation of the derived features are done for the Mg\,II\,k line formed at the quiet Sun and should not be assumed by default for any other line profile or line formation conditions without testing.
    
    Another application where autoencoders are often used is noise reduction. Figure~\ref{figure:examples} demonstrates that: the reconstructed line profiles (orange curves) look much smoother in comparison to the original line profiles (blue curves). Moreover, one can notice the small absorption feature on the left wing of the Mg\,II\,k line. This absorption is especially visible in Fig.~\ref{figure:examples}\textit{b,c}. This feature corresponds to the Mn\,I\,2795.64\,$\AA$ spectral line profile (see IRIS technical notes, \url{https://iris.lmsal.com/itn26/introduction.html}). It is again remarkable how the network captures it from the noisy spectroscopic data. Possibilities to reveal such weak spectral signatures from noisy data may also be utilized in the future.
    
    Finally, a relatively good reconstruction of the Mg\,II\,k line profile with the $n_{emb} = 4$ setting may have several practical applications, especially in situations where the data flow capabilities are limited. Compressing the data on deep space missions may result in more data samples transmitted to the ground via limited telemetry bandwidth. Low dimensionality of the embedding space allows one to perform a similarity search-based data retrieval~--- one can search data sets based on compact embedding instead of the entire data volume. Finally, line profile denoising, as we saw, may be useful for extracting weak signals in the continuum variations or weak spectroscopic signatures from noisy data.

\section{Conclusions}
\label{section:conclusion}

    In this paper we have studied the compact embedding of spectroscopic observations of the quiet Sun in the Mg\,II\,k line by NASA's IRIS satellite using the fully-connected autoencoders. The key outcomes of this study are as follows:
    \begin{itemize}
        \item It is possible to compress the data more than 27 times (i.e., to reduce the data dimensionality from 110 to 4) while having only a 4\,DN reconstruction error on average. The reconstruction error is somewhat comparable to the variations of the measurements in the line continuum observed for the line profiles;
        \item The average error of reconstruction of the MSE and even statistical moments of the line profiles decreases sharply for the $n_{emb} \leq{} 4$ and barely drops for $n_{emb} > 4$. Reconstruction of the odd statistical line moments does not demonstrate such dependence on the dimensionality of the embedding layer (except for the $\mu{}_{1}$ statistical moment);
        \item Some occasional improvements of the autoencoder training were observed for $n_{emb} > 4$, e.g., reducing MSE by $\sim$20\% for $n_{emb} = 8$ and by 55\% for $n_{emb} = 14$. However, these results were observed only twice among more than 100 runs, indicating that obtaining a better embedding requires, probably, longer training times and another training/architecture strategy;
        \item The features learned for the $n_{emb} = 4$ case can be supported with an intuitively-understandable interpretation if variations in the embedding space are considered. Three of the identified features were found to be mostly responsible for defining the line width, asymmetry, and line dip formation respectively.
    \end{itemize}
    
    In the presented study we considered only the Mg\,II\,k line profiles observed at a quiet Sun disc center region with a certain configuration of the observing instrument (exposure and resolution). While we expect that the techniques of compression of line profiles may be extended to other line profiles, this will require a detailed assessment. The presented work considered a particular architecture of the neural network, a particular way of varying the embedding layer dimensionality, and a certain methodology to train the network. Using other architectures such as the convolutional neural networks and variational autoencoders \cite{Kingma14}, as well as other training strategies, may help to better compress the line profiles and will be investigated later. Nevertheless, the work illustrates the potential of line-profile compact embedding for both operational and scientific purposes.

\section*{Acknowledgment}
VMS is supported by NSF FDSS grant 1936361 and NSF grant 1835958. EI is supported by RSF grant 20-72-00106. The research is partially supported by NASA grants NNX12AD05A, 80NSSC20K0302, and NSF EarthCube grant 1639683. IRIS is a NASA small explorer mission developed and operated by LMSAL with mission operations executed at NASA Ames Research center and major contributions to downlink communications funded by ESA and the Norwegian Space Centre.


\begin{thebibliography}{00}

\bibitem{DePontieu2014} B. De Pontieu, A. Title, J. Lemen, G. Kushner, D. Akin, et al., ``The Interface Region Imaging Spectrograph (IRIS)'', Solar Physics, vol. 289, pp. 2733-2779, July 2014.

\bibitem{Leenaarts2013} J. Leenaarts, T. Pereira, M. Carlsson, H. Uitenbroek, and B. De Pontieu, ``The Formation of IRIS Diagnostics. II. The Formation of the Mg II h\&k Lines in the Solar Atmosphere'', The Astrophysical Journal, vol. 772, p. 90, August 2013.

\bibitem{Sadykov2021} V. Sadykov, I. Kitiashvili, A. Kosovichev, and A. Wray, ``Connecting Atmospheric Properties and Synthetic Emission of Shock Waves Using 3D RMHD Simulations of Quiet Sun'', The Astrophysical Journal, in production, 2021.

\bibitem{SainzDalda2019} A. Sainz Dalda, J. de la Cruz Rodriguez, B. De Pontiey, and M. Go{\v{s}}i{\'c}, ``Recovering Thermodynamics from Spectral Profiles observed by IRIS: A Machine and Deep Learning Approach'', The Astrophysical Journal Letters, vol. 875, L18, April 2019.

\bibitem{Panos2018} B. Panos, L. Kleint, C. Huwyler, S. Krucker, M. Melchior, D. Ullmann, and S. Voloshynovskiy, ``Identifying Typical Mg II Flare Spectra Using Machine Learning'', The Astrophysical Journal, vol. 861, p. 62, July 2018.

\bibitem{Nita2020} Nita, G., Georgoulis, M., Kitiashvili, I., Sadykov, V. et al., ``Machine Learning in Heliophysics and Space Weather Forecasting: A White Paper of Findings and Recommendations'', ArXiv 2006.12224, 2020.

\bibitem{Panos2021} Panos, B., Kleint, L., Voloshynovskiy, S., ``Exploring mutual information between IRIS spectral lines. I. Correlations between spectral lines during solar flares and within the quiet Sun'', ArXiv 2104.12161, 2021.

\bibitem{Paszke2017} A. Paszke, S. Gross, S. Chintala, G. Chanan, et al., ``Automatic differentiation in PyTorch'', 31st Conference on Neural Information Processing Systems (NISP 2017), Long Beach, CA, USA, 2017

\bibitem{Kingma2015} D. Kingma and J. Ba, ``Adam: {A} Method for Stochastic Optimization'', 3rd International Conference on Learning Representations ({ICLR} 2015), Conference Track Proceedings, San Diego, CA, USA, May 7-9, 2015.

\bibitem{Sadykov2015} V. Sadykov, S. Vargas Dominguez, A. Kosovichev, I. Sharykin, A. Struminsky, and I. Zimovets, ``Properties of Chromospheric Evaporation and Plasma Dynamics of a Solar Flare from IRIS Observations'', The Astrophysical Journal, vol. 805, p. 167, June 2015.

\bibitem{Kingma14} D.P. Kingma and M. Welling, ``Auto-Encoding Variational Bayes'', Conference Track Proceedings, 2nd International Conference on Learning Representations, ICLR 2014, Banff, AB, Canada, April 14-16, 2014.

\end{thebibliography}
\end{document}